\documentclass[10pt,journal,twocolumn,twoside]{IEEEtran}
\usepackage{epsfig,amsfonts,subfigure,color,amsbsy}
\usepackage[nolist]{acronym}
\usepackage{graphicx,cite,amssymb,amsthm,bm}
\usepackage{amsmath}
\usepackage{mathrsfs}
\usepackage{bm,bbm}
\usepackage{multirow}
\usepackage{bigstrut}
\usepackage{psfrag}
\usepackage{stackengine}
\usepackage{xcolor}
\usepackage{mathtools}
\usepackage{tabularx,array}
\usepackage{mathbbol}
\usepackage{lipsum}
\usepackage[sans]{dsfont}
\mathtoolsset{showonlyrefs}
\usepackage{array, makecell}
\usepackage{float}
\usepackage{balance}
\usepackage{verbatim}
\usepackage{mathrsfs}
\usepackage{scalerel}
\usepackage{nicematrix}
\usepackage{bigints}
\newcommand{\pmata}[2]{\bm{\Pi}_{\scaleto{\mathsf{A},#2}{5pt}}^{\scaleto{(#1)}{5pt}}}
\newcommand{\pmatc}[2]{\bm{\Pi}_{\scaleto{\mathsf{C},#2}{5pt}}^{\scaleto{(#1)}{5pt}}}
\newcommand{\sdot}[1]{\scaleto{#1}{9.5pt}}
\newcommand{\ch}[1]{\scaleto{\textsf{#1}}{6.5pt}}
\newcommand{\eps}[1]{\epsilon_{\scaleto{\textsf{#1}}{4pt}}}
\newcommand{\gam}[1]{\gamma_{\scaleto{\textsf{#1}}{4pt}}}
\newcommand{\ent}[1]{\mathtt{h}_{\scaleto{\textsf{#1}}{4pt}}}
\newcommand{\prb}[1]{p_{\scaleto{\textsf{#1}}{4pt}}}

\newcommand{\inR}{R_{\scaleto{\mathsf{I}}{3.5pt}}}

\newcommand{\dmR}{R_{\scaleto{\mathsf{O}}{3.5pt}}}
\newcommand{\Hb}{\mathsf{H}_{\scaleto{\mathsf{b}}{4pt}}}

\newcommand{\liftfac}{\ell}
\newcommand{\hw}{w_{\scaleto{\mathsf{H}}{3.5pt}}}

\newcommand{\pgraph}{\mathcal{P}}
\newcommand{\bgraph}{\mathcal{G}}

\newcommand{\vn}{\mathsf{v}}
\newcommand{\cn}{\mathsf{c}}
\newcommand{\transposed}{{\scaleto{\mathsf{T}}{3.5pt}}}

\newcommand{\vB}{\bm{B}}
\newcommand{\vBSC}{\bm{B}_{\scaleto{\mathsf{SC}}{3.5pt}}}

\newcommand{\vX}{\bm{X}}

\newcommand{\edges}{\mathscr{E}}

\newcommand{\cV}{\mathcal{V}}
\newcommand{\cC}{\mathcal{C}}

\IEEEoverridecommandlockouts
\begin{document}

\makeatletter

\newtheorem{exampleinner}{Example}[section]
\renewcommand{\theexampleinner}{\arabic{exampleinner}} % only number, no section prefix
\newenvironment{example}{\begin{exampleinner}}{\end{exampleinner}}

%%%%%%%%%%%%%%%%%%%%%%%%%%%%%%%%%%%%%%%%%%%%%%%%%%%%%%%%%%%%%%%%%%%%%%%%%%%%%%%%%%%%%%%%%%%%%%%%%%%%%%%%%%

\title{Rate-Adaptive Spatially Coupled MacKay-Neal Codes with Thresholds Close to Capacity}
\author{Ayman Zahr, \IEEEmembership{Student Member, IEEE}, and Gianluigi Liva, \IEEEmembership{Senior Member, IEEE}
	\thanks{ 
		A. Zahr is with the Institute for Communications Engineering, Technical University of Munich, Munich, Germany, and with the Institute of Communications and Navigation, German Aerospace Center (DLR), Wessling, Germany (email: ayman.zahr@dlr.de).
		G. Liva is with the Institute of Communications and Navigation, German Aerospace Center (DLR), Wessling, Germany (email: gianluigi.liva@dlr.de).
	}
	\thanks{
		The authors acknowledge the financial support by the Federal Ministry of Research, Technology and Space of Germany in the programme of ``Souver\"an. Digital. Vernetzt.'' Joint
		project 6G-RIC, project identification number: 16KISK022. 
	}
}

\maketitle
\thispagestyle{empty}
\pagestyle{empty}
\IEEEoverridecommandlockouts

%%%%%%%%%%%%%%%%%%%%%%%%%%%%%%%%%%%%%%%%%%%%%%%%%%%%%%%%%%%%%%%%%%%%%%%%%%%%%%%%%%%%%%%%%%%%%%%%%%%%%%%%%%

%%%%%%%%%%%%%%%%%%%%%%%%%%%%%%%%%%%%%%%%%%%%%%%%%%%%%%%%%%%%%%%%%%%%%%%%%%%%%%%%%%%%%%%%%%%%%%%%%%%%%%%%%%

\begin{abstract}
	We analyze by density evolution the asymptotic performance of rate-adaptive  MacKay-Neal (MN) code ensembles, where the inner code is a protograph spatially coupled (SC) low-density parity-check code. By resorting to a suitably-defined parallel channel model, we compute belief propagation decoding thresholds, showing that SC MN code ensembles can perform within $0.15$ dB from the binary-input additive white Gaussian noise capacity over the full $[0,1]$ rate range.
\end{abstract}

%%%%%%%%%%%%%%%%%%%%%%%%%%%%%%%%%%%%%%%%%%%%%%%%%%%%%%%%%%%%%%%%%%%%%%%%%%%%%%%%%%%%%%%%%%%%%%%%%%%%%%%%%%

\pagenumbering{arabic}

%%%%%%%%%%%%%%%%%%%%%%%%%%%%%%%%%%%%%%%%%%%%%%%%%%%%%%%%%%%%%%%%%%%%%%%%%%%%%%%%%%%%%%%%%%%%%%%%%%%%%%%%%%

\begin{acronym}
	\acro{BEC}{binary erasure channel}
	\acro{BP}{belief propagation}
	\acro{DE}{density evolution}
	\acro{LDPC}{low-density parity-check}
	\acro{SC-LDPC}{spatially coupled low-density parity-check}
	\acro{ML}{maximum likelihood}
	\acro{MAP}{maximum a posteriori}
	\acro{r.v.}{random variable}
	\acro{PEP}{pairwise error probability}
	\acro{BP}{belief propagation}
	\acro{BPSK}{binary phase shift keying}
	\acro{BSC}{binary symmetric channel}
	\acro{AWGN}{additive white Gaussian noise}
	\acro{OOK}{on-off keying}
	\acro{DM}{distribution matcher}
	\acro{p.m.f.}{probability mass function}
	\acro{p.d.f.}{probability density function}
	\acro{i.i.d.}{independent and identically-distributed}
	\acro{CC}{constant composition}
	\acro{LEO}{low earth orbit}
	\acro{biAWGN}{binary-input additive white Gaussian noise}
	\acro{PAM}{pulse amplitude modulation}
	\acro{SNR}{signal-to-noise ratio}
	\acro{EXIT}{extrinsic information transfer}
	\acro{PEXIT}{protograph extrinsic information transfer}
	\acro{VN}{variable node}
	\acro{CN}{check node}
	\acro{FER}{frame error rate}
	\acro{MET}{multi-edge type}
	\acro{MN}{MacKay-Neal}
	\acro{RA}{repeat-accumulate}
	\acro{NS}{non-systematic}
	\acro{SC}{spatially coupled}
	\acro{LLR}{log-likelihood ratio}
	\acro{MM}{mismatched}
	\acro{EPC}{\emph{equivalent parallel channel}}
	\acro{WCL}{worst-case loss}
	\acro{UB}{union bound}
	\acro{TUB}{truncated union bound}
	\acro{PEG}{progressive edge growth}
	\acro{PAS}{probabilistic amplitude shaping}
	\acro{CCDM}{constant composition distribution matcher}
	\acro{MI}{mutual information}
	\acro{BMS}{binary-input memoryless symmetric}
\end{acronym}

%%%%%%%%%%%%%%%%%%%%%%%%%%%%%%%%%%%%%%%%%%%%%%%%%%%%%%%%%%%%%%%%%%%%%%%%%%%%%%%%%%%%%%%%%%%%%%%%%%%%%%%%%%
%%%%%%%%%%%%%%%%%%%%%%%%%%%%%%%%%%%%%%%%%%%%%%%%%%%%%%%%%%%%%%%%%%%%%%%%%%%%%%%%%%%%%%%%%%%%%%%%%%%%%%%%%%
%%%%%%%%%%%%%%%%%%%%%%%%%%%%%%%%%%%%%%%%%%%%%%%%%%%%%%%%%%%%%%%%%%%%%%%%%%%%%%%%%%%%%%%%%%%%%%%%%%%%%%%%%%

\section{Introduction}\label{sec:intro}

\IEEEPARstart{M}{odern} high‐throughput communication systems demand flexible, rate‐adaptive error‐correction solutions that avoid costly hardware reconfiguration. To address this need, a class of protograph-based \ac{MN} codes was introduced and analyzed in \cite{zahr2024rate}. The code structure closely follows the one introduced in \cite{Mac99}. In particular, an outer nonlinear encoder is concatenated with an inner nonsystematic \ac{LDPC} \cite{Gal63} code encoder. The outer nonlinear encoder, named \ac{DM} \cite{Schulte2016}, maps the input message onto a fixed-length sequence with a prescribed empirical distribution via a binary \ac{CC} code. The sequence is then input to the nonsystematic \ac{LDPC} encoder, whose output is transmitted over the channel. At the decoder side, \ac{BP} decoding is performed over the bipartite graph of the inner \ac{LDPC} code, where the \acp{VN} associated with the encoder input are fed with prior information derived from the marginal distribution of the outer \ac{CC} code. By changing the distribution defined by the \ac{DM}, a large range of code rates can be obtained without modifying (i.e., puncturing or shortening) the inner code. The construction of \cite{zahr2024rate} relies on inner protograph-based \ac{LDPC} codes. 
The optimization of the inner code protograph allows to operate, al large blocklength, within  $1$ dB from the \ac{biAWGN} channel capacity. To achieve a robust performance across the various rates, the design of \cite{zahr2024rate} uses \ac{DE} \cite{RU01a} analysis, seeking protographs for which the maximum gap between the \ac{BP} thresholds and the Shannon limit is minimized over a set of rates. 

In this paper, we explore the performance of protograph-based \ac{MN} code ensembles where the inner code is a protograph \ac{SC} \ac{LDPC} code \cite{FZ+99,Lentmaier2010,KRU+11,MLC+15}. This choice is motivated by the capacity approaching properties of \ac{SC} \ac{LDPC} code ensembles \cite{Lentmaier2010,KRU+11}, and by their universality of their performance w.r.t. a large class of channels \cite{Kudekar2013}---a property that may be instrumental to close the gap to the Shannon limit for the resulting \ac{SC} \ac{MN} construction, at all rates. A different class of \ac{SC} \ac{MN} codes was analyzed via \ac{DE} in \cite{Kasai2011}. A key difference between the class of codes studied in \cite{Kasai2011} and the one addressed in this paper stems from the absence, in \cite{Kasai2011}, of the concatenation with an outer nonlinear code, i.e., the analysis of \cite{Kasai2011} considers \emph{linear} \ac{SC} \ac{MN} codes that constitute a subclass of multi-edge type \ac{SC} \ac{LDPC} codes. 

We analyze \ac{SC} \ac{MN} codes obtained by concatenating an outer nonlinear (\ac{CC}) code with an inner regular protograph \ac{SC} \ac{LDPC} code. The analysis is carried out in the asymptotic regime of large lifting factors, and long \ac{SC} chains, by means of \ac{DE} analysis. The analysis relies on the \ac{EPC} model introduced in \cite{zahr2024rate} to circumvent the challenges posed by the nonlinear code structure. We show that \ac{SC} \ac{MN} codes can operate within $0.15$ dB from the Shannon limit for the \ac{biAWGN} channel over the full rate range $[0,1]$, with a single inner \ac{SC} \ac{LDPC}, and with rate adaptation achieved by tuning the outer \ac{DM} parameter.

%%%%%%%%%%%%%%%%%%%%%%%%%%%%%%%%%%%%%%%%%%%%%%%%%%%%%%%%%%%%%%%%%%%%%%%%%%%%%%%%%%%%%%%%%%%%%%%%%%%%%%%%%%

\section{Preliminaries}\label{sec:prel}
In this paper, \acp{r.v.} are denoted by uppercase letters, and their realization by lowercase letters. The \ac{p.d.f.} of a \ac{r.v.} $X$ is denoted by $p(x)$. The binary entropy function $\Hb(\omega)$ is $\Hb(\omega)=-\omega \log_2 \omega -(1-\omega)\log_2(1-\omega)$, for $0 < \omega < 1$ and $\Hb(0)=\Hb(1)=0$. Vectors are treated as row vectors and denoted by bold letters, e.g., $\bm{x}$, while matrices are denoted by uppercase bold letters, e.g., $\vX$. %We use $\fieldtwo$ to represent the binary finite field. 
The Hamming weight of $\bm{x}$ is denoted by $\hw(\bm{x})$. We consider transmission over the \ac{biAWGN} channel, with $Y = (-1)^X + N$, where $X\in\{0,1\}$, $N\sim\mathcal{N}(0,\sigma^2)$ is the additive white Gaussian noise term, and $Y$ is the channel output. The \ac{SNR} is $E_s/N_0 = 1/(2\sigma^2)$, where $E_s$ is the energy per symbol and $N_0$ is the single-sided noise power spectral density.

\subsection{Protograph-based (Spatially-Coupled) LDPC Ensembles}\label{sec:prel:protographs}

For the \ac{MN} code construction, we rely on protograph-based \ac{LDPC} and \ac{SC} {LDPC} code ensembles. In particular, we focus on two ensembles: the $(d_v, d_c)$ regular block \ac{LDPC} code ensemble, where $d_v$ and $d_c$ denote the \ac{VN} and \ac{CN} degrees, and the corresponding \ac{SC} code ensemble. 

A protograph $\pgraph = (\cV , \cC, \edges)$ is a small bipartite graph  consisting of a set $\cV$ of  $\mathtt{N}$ \acp{VN}, a set $\cC$ of $\mathtt{M}$ \acp{CN}, and a set $\edges$ of $e$ edges \cite{Tho03}. \acp{VN} in the protograph are numbered from $0$ to $\mathtt{N}-1$. Similarly, protograph \acp{CN} are numbered from $0$ to $\mathtt{M}-1$. Each \ac{VN}/\ac{CN}/edge in a protograph defines a \ac{VN}/\ac{CN}/edge type.  
The bipartite graph $\bgraph$ of an \ac{LDPC} code can be derived by lifting the protograph. In particular, the protograph is copied ${\liftfac}$ times (where ${\liftfac}$ is referred to as the \emph{lifting factor}), and the edges of the protograph copies are permuted under the following constraint: if an edge connects a type-$j$ \ac{VN} to a type-$i$ \ac{CN} in $\pgraph$, after permutation the edge should connect one of the ${\liftfac}$ type-$j$ \ac{VN} copies with one of the ${\liftfac}$ type-$i$ \ac{CN} copies in $\bgraph$. We denote by $\vn_0,\vn_1,\ldots$  \acp{VN} in $\bgraph$, and by $\cn_0,\cn_1,\ldots$ the \acp{CN} in $\bgraph$.
The lifted graph $\bgraph$ defines the parity-check matrix of an \ac{LDPC} code.  The base matrix of a protograph is an $\mathtt{M} \times \mathtt{N}$ matrix $\vB= [b_{i,j}]$ where $b_{i,j}$ is the number of edges that connect \ac{VN} $j$ to \ac{CN} $i$ in $\pgraph$. We will make use of \ac{LDPC} codes with \emph{punctured} (or \emph{state}) \acp{VN}. A punctured \ac{VN} is associated with a codeword bit that is not transmitted through the communication channel. We will assume that all the \acp{VN} of a given type are either punctured or they are not, i.e., puncturing is defined at protograph level.

\begin{figure}
	\centering 
	\includegraphics[width=0.65\columnwidth]{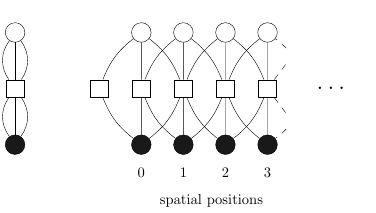}
	\caption{$(3, 6)$ protograph \ac{SC} LDPC code ensemble from Example~\ref{ex:base} (left) and of its SC counterpart (right). Dark variable nodes denote punctured nodes.}
	\label{fig:protograph_LDPC}
\end{figure}

Protographs can be used to define both block and \ac{SC} \ac{LDPC} code ensembles. In this paper, we focus on regular block code ensembles defined by a base matrix in the form 
\begin{equation}\label{eq:baseblock}
	\vB = \begin{pmatrix}
		d & d
	\end{pmatrix}
\end{equation}
where the \ac{VN} degree is $d_v=d$ and the \ac{CN} degree is $d_c = 2d$. An example of a protograph defined by a base matrix in the form \eqref{eq:baseblock} with $d=3$ is depicted in Fig.~\ref{fig:protograph_LDPC} (left side). By convention, the punctured \ac{VN} (dark circle) is associated with the first column of $\vB$.
We consider \ac{SC} \ac{LDPC} code ensembles defined by protographs in the form
\begin{align}
	\vBSC &= \left(
	\begin{array}{cccc}
		\vB_{0} &          &          \\
		\vB_{1} &  \vB_{0} &          \\
		\sdot{\vdots}  &  \vB_{1} & \sdot{\ddots}  \\
		\vB_{d-1} &  \sdot{\vdots}  & \sdot{\ddots} \\
		&  \vB_{d-1} &        \\
		&          & \sdot{\ddots}
	\end{array}
	\right)
	\label{eq:base_matrix_SC}
\end{align}
where $\vB_{0} = \vB_{1} = \ldots = \vB_{d-1} = \begin{pmatrix}
	1 & 1
\end{pmatrix}$. The number of column blocks (spatial positions) is denoted by $\mathtt{L}$. An example of a protograph defined by a base matrix in the form \eqref{eq:base_matrix_SC} with $d=3$ is given in Fig.~\ref{fig:protograph_LDPC} (right side). As before, by convention, we associate the punctured \acp{VN} to even column indexes, as illustrated by the following example.

\newcommand{\blk}{\Block[fill=black!15,rounded-corners]{6-1}{}}
\newcommand{\wth}[1]{\textcolor{black}{#1}}
\begin{example}\label{ex:base}
	The base matrix of \ac{SC} \ac{LDPC} code ensemble defined according to \eqref{eq:base_matrix_SC}, for $d=3$, is
	\begin{equation}\label{eq:Bsc3}
		\vBSC = \begin{pNiceMatrix}[margin]
			\blk\wth{1} &  1 & \blk\wth{}   &   & \blk\wth{}  &   & \wth{} \\
			\wth{1} &  1 &     \wth{1}   & 1  &     \wth{}  &   &     \wth{} \\
			\wth{1} &  1 &     \wth{1}   & 1  &     \wth{1}  & 1  &     \wth{} \\
			\wth{} &   &     \wth{1}   & 1  &     \wth{1}  & 1  &     \sdot{\ddots} \\
			\wth{} &   &     \wth{}   &   &     \wth{1}  & 1  &     \sdot{\ddots} \\
			\wth{} &   &     \wth{}   &   &     \wth{}  &   &     \sdot{\ddots} \\
		\end{pNiceMatrix}
	\end{equation}
	where the columns with index $0, 2, 4, \ldots$ (marked in gray) are associated to punctured \acp{VN}.
\end{example}

Since all nonzero entries in the base matrix \eqref{eq:base_matrix_SC} are set to $1$, lifting is performed by replacing each nonzero entry with an $\ell \times \ell$ permutation matrix, and each zero entry with an $\ell \times \ell$ zero matrix. The resulting parity check matrix is in the form
\begin{equation}\label{eq:Hsc}
	\bm{H} = \begin{pNiceMatrix}[margin]
		\blk\wth{\pmata{0}{0}} &  \pmatc{0}{0} & \blk\wth{}   &   &   \\
		\wth{\pmata{0}{1}} &  \pmatc{0}{1} &     \pmata{1}{0}   & \pmatc{1}{0}  &      \\
		\sdot{\vdots} &  \sdot{\vdots} &     \pmata{1}{1}   & \pmatc{1}{1}  &     \sdot{\ddots}  \\
		\wth{\pmata{0}{d-1}} &  \pmatc{0}{d-1} &     \sdot{\vdots}   & \sdot{\vdots}  &     \sdot{\ddots}  \\
		&  &     \pmata{1}{d-1}   & \pmatc{1}{d-1}  &        \\
		&  &        &  &     \sdot{\ddots}   
	\end{pNiceMatrix}
\end{equation}
where, following the visual convention introduced in Example~\ref{ex:base}, columns marked in gray refer to punctured \acp{VN}
(the meaning of the subscripts $\mathsf{A}$ and $\mathsf{C}$, associated with punctured and unpunctured \acp{VN}, will be explained in Section~\ref{sec:analysis}).

\begin{figure*}
	\begin{center}
		\includegraphics[width=0.85\textwidth]{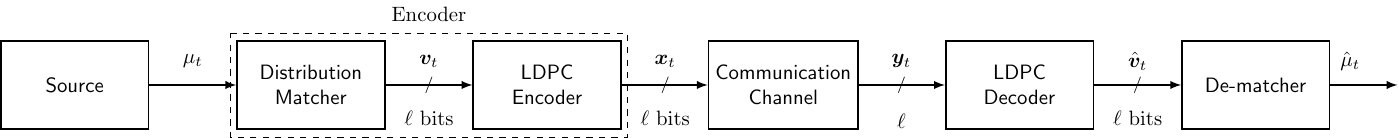}
		\vspace{-2mm}
		\caption{System model, where a \ac{SC} \ac{MN} code is used to communicate over the \ac{biAWGN} channel (communication channel).}\label{fig:model}
	\end{center}
	\vspace{-3mm}
\end{figure*}

%%%%%%%%%%%%%%%%%%%%%%%%%%%%%%%%%%%%%%%%%%%%%%%%%%%%%%%%%%%%%%%%%%%%%%%%%%%%%%%%%%%%%%%%%%%%%%%%%%%%%%%%%%

\section{Spatially Coupled MacKay-Neal Codes}\label{sec:model}

As in \cite{zahr2024rate}, we construct a protograph-based \ac{MN} code as the concatenation of an outer \ac{DM} (nonlinear \ac{CC} code) and an inner protograph-based \ac{LDPC} code with nonsystematic encoder. Differently from \cite{zahr2024rate}, where the inner code is a block \ac{LDPC} code, we use here a protograph-based \ac{SC} \ac{LDPC} code whose base matrix complies with \eqref{eq:base_matrix_SC}. To illustrate the code structure, we focus on the encoder (Fig.~\ref{fig:model}). We assume a source that emits, at discrete times,  a stream of messages $\mu_0, \mu_1,\ldots$ with $\mu_t \in \{1,2,\ldots,M\}$. The generic message $\mu_t$ is encoded by the \ac{DM} into a binary $\ell$-tuple $\bm{v}_t$, where $\bm{v}_t$ has a prescribed (constant) Hamming weight $\hw(\bm{v}_t)=\omega\ell$, for some $\omega \in \{0, 1/\ell, 2/\ell, \ldots, 1\}$. We refer to the \ac{DM} parameter $\omega$ as the fractional Hamming weight of $\bm{v}_t$. The pair $(\omega$, $\ell)$ defines the outer \ac{CC} code. The \ac{CC} outer codewords $\bm{v}_0, \bm{v}_1, \ldots$ are fed into the inner nonsystematic \ac{SC} \ac{LDPC} encoder, producing a sequence of $\ell$-tuples $\bm{x}_0, \bm{x}_1, \ldots$ that are sent over the communication channel. According to the parity-check matrix structure of \eqref{eq:Hsc}, encoding is performed as
\begin{equation}
	\bm{x}_t^\transposed = \left[\pmatc{t}{1}\right]^{-1}\left(\sum_{i=0}^{d-1} \pmata{t-i}{i}\bm{v}_{t-i}^\transposed + \sum_{i=1}^{d-1} \pmatc{t-i}{i}\bm{x}_{t-i}^\transposed\right)
\end{equation}	
where $\bm{v}_t = \bm{0}$ and $\bm{x}_t = \bm{0}$ for $t<0$.

We assume zero-tail termination of the \ac{SC} \ac{LDPC} code. Following \cite{Lentmaier2010,MLC+15}, the encoder stops encoding the \ac{DM} output sequence at time $t = \mathtt{L} - d$. The last $d-1$ \ac{SC} \ac{LDPC} encoder inputs $\bm{v}_{\mathtt{L}-d+1}, \bm{v}_{\mathtt{L}-d+2}, \ldots, \bm{v}_{\mathtt{L}-1}$ are computed to drive the encoder to the zero state (zero syndrome) at time $\mathtt{L}-1$. In the proposed construction, the encoder inputs $\bm{v}_{\mathtt{L}-d+1}, \bm{v}_{\mathtt{L}-d+2}, \ldots, \bm{v}_{\mathtt{L}-1}$ are not punctured, i.e., they are transmitted over the communication channel along with the corresponding encoder outputs $\bm{x}_{\mathtt{L}-d+1}, \bm{x}_{\mathtt{L}-d+2}, \ldots, \bm{x}_{\mathtt{L}-1}$. The inner code rate is hence
\begin{equation}\label{eq:Ri}
	\inR = \frac{\mathtt{L} - (d - 1)}{\mathtt{L} - (d - 1) + 2(d-1)} = \frac{\mathtt{L} - (d - 1)}{\mathtt{L} + (d - 1)}.
\end{equation}
The inner code rate tends to one as $\mathtt{L}\rightarrow \infty$. The rate of the outer code (\ac{DM}) is \cite{zahr2024rate} 
\begin{equation}\label{eq:Ro}
	\dmR = \frac{1}{\ell}\log_2 {\ell \choose \omega\ell} 
\end{equation}
and it converges to $\Hb(\omega)$ for large $\ell$. The overall rate is hence $R = \dmR \inR$, and it approaches $\Hb(\omega)$ in the asymptotic regime where both $\mathtt{L}$ and $\ell$ tend to infinity. Note that, by selecting $\omega \in [0,1/2]$, it is possible to finely span the full code rate region $R\in [0,1]$, enabling flexible rate adaptation.

\ac{BP} decoding takes place on the graph of the inner \ac{LDPC} code. For unpunctured \acp{VN}, the input is given by their corresponding channel \acp{LLR}, i.e., $L = \ln [{p(y|0)}/{p(y|1)}]$.
Punctured \acp{VN} are initialized with the prior obtained from the marginal distribution of the \ac{DM} output \cite{Mac99,bocherer2015bandwidth,zahr2024rate}, i.e., 
$L = \ln [(1-\omega)/{\omega}]$.
The hard decisions at the \ac{LDPC} decoder output, $\hat{\bm{v}}_0,\hat{\bm{v}}_1, \ldots$,  are passed to the de-matcher, producing the sequence of message estimates $\hat{\mu}_0,\hat{\mu}_1, \ldots$.

\begin{figure*}
	\begin{center}
		\includegraphics[width=0.85\textwidth]{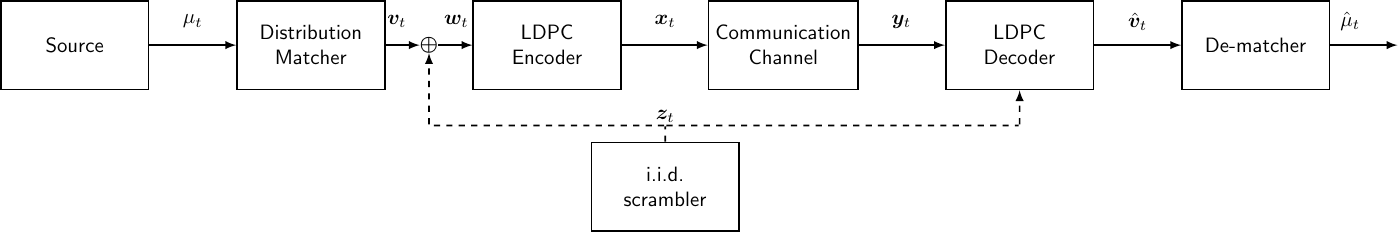}
		\vspace{-2mm}
		\caption{Modified \ac{SC} \ac{MN} scheme, with the introduction of the bit scrambler.}\label{fig:equiv}
	\end{center}
	\vspace{-3mm}
\end{figure*}

%%%%%%%%%%%%%%%%%%%%%%%%%%%%%%%%%%%%%%%%%%%%%%%%%%%%%%%%%%%%%%%%%%%%%%%%%%%%%%%%%%%%%%%%%%%%%%%%%%%%%%%%%%

\section{Density Evolution Analysis}\label{sec:analysis}

\begin{figure}
	\centering
	{
		\includegraphics[width=0.9\columnwidth]{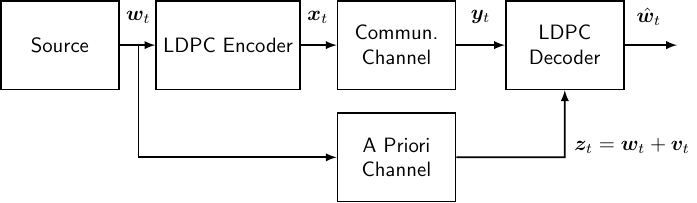}
	}
	\vspace{-1mm}
	\caption{Equivalent parallel channel model.}
	\label{fig:equiv2}
	\vspace{-3mm}
\end{figure}

As observed in \cite{zahr2024rate}, the analysis of the \ac{MN} code construction can be challenging, due to the inherent nonlinearity of the code (caused by the concatenation with the outer, nonlinear, constant composition code).
The nonlinearity results in a (bit/block) error probability that is a function of the transmitted sequence. This aspect, in particular, is of impediment for the use of the all-zero codeword assumption used in the \ac{DE} analysis of the code ensemble. In \cite{zahr2024rate}, the problem was circumvented by observing that the introduction of a random \ac{i.i.d.} bit scrambler, flipping the bits at the output of the \ac{DM} independently and with probability $1/2$ (with scrambling sequence made available at the decoder), does not modify the error probability of the scheme over symmetric communication channels. Hence, the performance analysis of the scheme of Fig.~\ref{fig:model} (without the scrambler) is equivalent to the one of the scheme of Fig.~\ref{fig:equiv}. By interpreting the scrambling sequence $\bm{z}_t$ as the output of a binary memoryless symmetric source, and the \ac{DM} output $\bm{v}_t$ as additive binary noise contribution, the analysis of the average error probability of the  \emph{nonlinear} \ac{MN} reduces to the analysis of the error probability of the inner \emph{nonpunctured, linear} \ac{LDPC} code (whose rate approaches $1/2$ as $\mathtt{L}\to \infty$), where the information bits are transmitted over an additive binary noise channel with input $\bm{w}_t$ and output $\bm{z}_t = \bm{w}_t + \bm{v_t}$, with marginal cross-over probability $\omega$, and where the parity bits $\bm{x}_t$ are transmitted over the \ac{biAWGN} channel, with channel output $\bm{y}_t$ \cite{fresia2010joint,GM22}. We refer to this model (illustrated in Fig.~\ref{fig:equiv2}) as the \ac{EPC} model. To emphasize their roles in the model, we refer to the \ac{biAWGN} channel as the \emph{communication channel} (channel-\ch{C}), and to  additive binary noise channel as \emph{a-priori channel} (channel-\ch{A}).\footnote{We are now able to clarify the notation used for the parity-check matrix structure of \eqref{eq:Hsc}. The subscript $\scaleto{\mathsf{A}}{5pt}$ used for the permutation matrices at even column positions emphasized that the corresponding \acp{VN} in the graph of the \ac{SC} \ac{LDPC} code are connected, in the \ac{EPC} model, to the a-priori channel. The subscript $\scaleto{\mathsf{C}}{5pt}$ used for the permutation matrices at odd column positions emphasized that the corresponding \acp{VN} in the graph of the \ac{SC} \ac{LDPC} code are connected, in the \ac{EPC} model, to the communication channel.}

\subsection{\ac{BP} Decoding Thresholds over Parallel Channels}
We initially analyze the performance of the \ac{SC} \ac{LDPC} code ensemble defined by \eqref{eq:base_matrix_SC} over various combinations of parallel channels, namely: (i) two parallel \acp{BEC}, where channel-\ch{A} has erasure probability $\eps{A}$ and channel-\ch{C} has erasure probability $\eps{C}$; (ii) two parallel \ac{biAWGN} channels, where channel-\ch{A} has $E_s/N_0 = \gam{A}$ and channel-\ch{C} has $E_s/N_0 = \gam{C}$; (iii) a \ac{BSC}-\ac{biAWGN} channel combination, where channel-\ch{A} is a \ac{BSC} with cross-over probability $\omega$ and channel-\ch{C} is a \ac{biAWGN} channel with \ac{SNR} $E_s/N_0$. The latter setting will serve as a proxy to the analysis of the \ac{BP} decoding thresholds of \ac{SC} \ac{MN} code ensembles over the \ac{biAWGN} channel (that is our original setting).
The analysis will be provided for the two ensembles defined by $d=3$ and $d=4$, that will be referred to as $(3,6)$ and $(4,8)$ regular \ac{SC} code ensembles, respectively.

With reference to the \ac{EPC} model of Fig.~\ref{fig:equiv2}, we parameterize the channels through their conditional entropies
\begin{align}
	\ent{A} = H(W|Z) \quad \text{and} \quad
	\ent{C} = H(X|Y). 
\end{align} 
Considering the asymptotic limit $\mathtt{L}\to \infty$, the unpunctured \ac{SC} \ac{LDPC} code rate is $R=1/2$. Note that, over the \ac{EPC} model, the Shannon limit for rate-$1/2$ codes yields the following upper bound on the sum of conditional entropies \cite{dedeoglu2015spatial}:
\begin{equation}\label{eq:limit}
	\ent{A} + \ent{C} \leq 1.
\end{equation}
In all cases, the \ac{BP} thresholds are computed via \ac{DE} by setting $\ent{A} = \ent{A}^\star$ and then by  determining what is the largest value of $\ent{C}$ ($\ent{A}^\star$) that yields a vanishing small bit error probability, in the large $\ell$ (lifting factor) limit. All results are obtained for a number of spatial positions $\mathtt{L}\to \infty$, hence removing the rate loss caused by the termination.

\subsubsection{BEC-BEC Parallel Channels}
The analysis is here reminiscent of the analysis introduced in \cite{dedeoglu2015spatial} to derive the \ac{BP} decoding thresholds of \ac{SC} root-\ac{LDPC} code ensembles. By setting $\eps{A} = \ent{A}$ and $\eps{C} = \ent{C}$,
the analysis follows by means of \ac{PEXIT} analysis \cite{LC07}, where the erasure probability associated to \acp{VN} connected to channel-\ch{A} is $\eps{A}$ and the erasure probability at the input of \acp{VN} connected to channel-\ch{C} is $\eps{C}$. The achievable threshold pairs ($\ent{A}^\star,\ent{C}^\star$) are depicted for $(3,6)$ and $(4,8)$ \ac{SC} code ensembles in Fig.~\ref{fig:thresholds_parallel} (first chart from the left), and are compared with the limit given by \eqref{eq:limit}. On the same chart, the achievable threshold pairs ($\ent{A}^\star,\ent{C}^\star$) for $(3,6)$ and $(4,8)$ regular protograph block ensembles, defined by a base matrix in the form \eqref{eq:baseblock}, are reported as reference. The results are close to the ones obtained in \cite{dedeoglu2015spatial} (albeit, for a different \ac{SC} root-\ac{LDPC} code ensemble). Similarly to what happens over single binary-input output symmetric memoryless channels, regular \ac{SC} \ac{LDPC} code ensembles allow to operate close to the limit, with a gap that is reduced by increasing the \ac{VN}/\ac{CN} degrees, largely outperforming their block code ensemble counterparts.

\subsubsection{biAWGN-biAWGN Parallel Channels}
\begin{figure*}[t]
	\begin{center}
		\includegraphics[width = 0.9\textwidth]{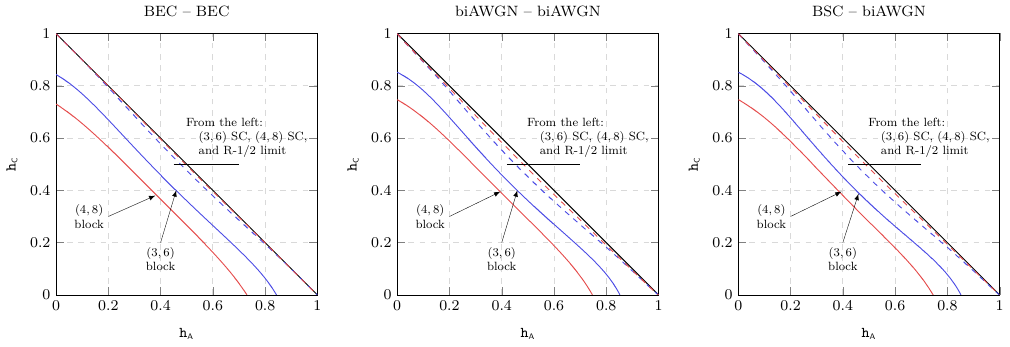}
		\caption{Thresholds regions for block \ac{MN} code ensembles (solid red/blue lines) and the corresponding \ac{SC} \ac{MN} code ensembles (dashed red/blue lines), compared to the rate-$1/2$ limit for parallel channel models (solid black line): \ac{BEC} -- \ac{BEC} (left), \ac{biAWGN} -- \ac{biAWGN} (center), and \ac{BSC} -- \ac{biAWGN} (right). }\label{fig:thresholds_parallel}
	\end{center}
\end{figure*}

By setting 
\begin{align}
	\gam{A}  =\frac{1}{8}\left[J^{-1}\left(1 -  \ent{A}\right)\right]^2 \quad \text{and} \quad
	\gam{C}  =\frac{1}{8}\left[J^{-1}\left(1 -  \ent{C}\right)\right]^2
\end{align}
where the $J(\cdot)$ defines the \ac{biAWGN} channel capacity \cite{ten01} 
\begin{equation}
	J(s)\!=\!1\!-\!\frac{1}{\sqrt{2\pi s^2}}\!\bigintssss_{-\infty}^{+\infty}\!\!\!\!\!\!\exp\left(\!-\frac{\left(u\!-\!s^2/2\right)^2}{2s^2}\right)\!\log_2\!\left(  1\!+\!e^{-u}\right)\mathrm{d} u \label{biawgnC}
\end{equation}
the analysis follows by means of the Gaussian approximation of \ac{DE}, i.e., via \ac{PEXIT} analysis \cite{LC07}. Here, the \ac{SNR} at the input of \acp{VN} connected to channel-\ch{A} is $\gam{A}$ and the \ac{SNR} at the input of \acp{VN} connected to  channel-\ch{C} is $\gam{C}$. The achievable threshold pairs ($\ent{A}^\star,\ent{C}^\star$) are depicted for $(3,6)$ and $(4,8)$ \ac{SC} code ensembles in Fig.~\ref{fig:thresholds_parallel} (middle chart), and are compared with the limit given by \eqref{eq:limit}. On the same chart, the achievable threshold pairs ($\ent{A}^\star,\ent{C}^\star$) for $(3,6)$ and $(4,8)$ regular protograph block ensembles, defined by a base matrix in the form \eqref{eq:baseblock}, are reported as reference. As for the \ac{BEC}-\ac{BEC} case, regular \ac{SC} \ac{LDPC} code ensembles allow to operate close to the limit, with a gap that is reduced by increasing the \ac{VN}/\ac{CN} degrees, largely outperforming their block code ensemble counterparts. The gap between the \ac{SC} \ac{LDPC} code ensembles and the limit is larger than the one observed in the \ac{BEC}-\ac{BEC} setting, especially when the  $\ent{A}, \ent{B}$ are close in value.

\subsubsection{BSC-biAWGN Parallel Channels}\label{subsubsec:BSCAWGN}
For the analysis over \ac{BSC}-\ac{biAWGN} parallel channels, we resort to quantized \ac{DE} \cite{Chung2001}, adapted to protographs. In particular, under the allzero codeword assumption, the conditional $L$-value distribution at the input of \acp{VN} connected to the a-priori channel is
\begin{equation}
	\prb{A}(L|W=0) = \omega \delta(L+\Delta) + (1-\omega) \delta(L-\Delta)
\end{equation}
where $\delta$ is the Dirac delta function, and 
$\Delta = \ln [(1-\omega)/{\omega}]$.
The conditional $L$-value distribution at the input of \acp{VN} connected to the communication channel is
\begin{equation}
	\prb{C}(L|X=0) = \mathcal{N}\left(4\gam{C}, 8\gam{C}\right)
\end{equation}
where $\gam{C}$ is the \ac{SNR} of the \ac{biAWGN} communication channel.
The conditional entropies of the a-priori channel and of the communication channel are $\ent{A} = \Hb(\omega)$ and $\ent{C}=1-J\left(\sqrt{8\gam{C}}\right)$, respectively. The achievable threshold pairs ($\ent{A}^\star,\ent{C}^\star$) are depicted for $(3,6)$ and $(4,8)$ \ac{SC} code ensembles in Fig.~\ref{fig:thresholds_parallel} (third chart from the left), and are compared with the limit given by \eqref{eq:limit}. On the same chart, the achievable threshold pairs ($\ent{A}^\star,\ent{C}^\star$) for $(3,6)$ and $(4,8)$ regular protograph block ensembles, defined by a base matrix in the form \eqref{eq:baseblock}, are reported as reference. The results are extremely close to the ones obtained for the \ac{biAWGN}-\ac{biAWGN} case, suggesting that modeling the a-priori \ac{BSC} as a \ac{biAWGN} channel can provide reliable threshold estimates. Considering the extreme simplicity of the \ac{PEXIT} analysis---which tracks only a single parameter for each protograph edge---this result is particularly important. In contrast, protograph quantized \ac{DE} tracks the evolution of a full message distribution per edge (for threshold calculations, we used $1023$ message quantization levels).

\subsection{BP Decoding Thresholds of \ac{SC} \ac{MN} Code Ensembles}
\begin{figure}[t]
	\begin{center}
		\includegraphics[width = 0.95\columnwidth]{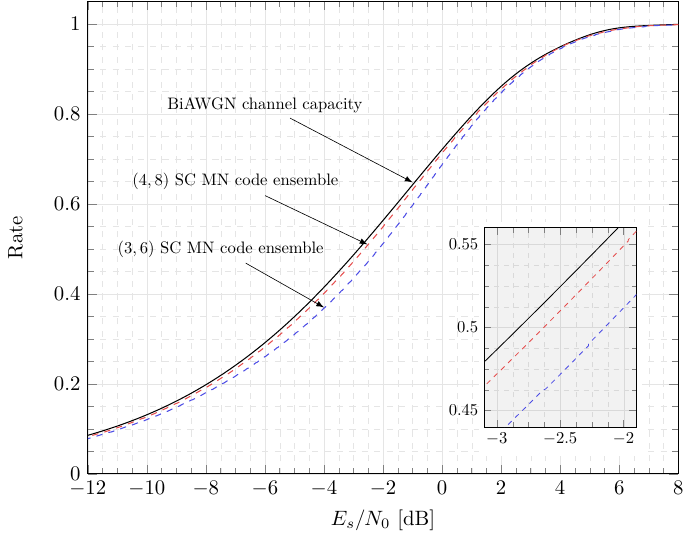}
		\caption{\ac{BP} thresholds of $(3,6)$ and $(4,8)$ \ac{SC} \ac{MN}  code ensembles computed via quantized \ac{DE}, and compared with the \ac{biAWGN} channel capacity.}\label{fig:thresholds_cap}
	\end{center}
\end{figure}

For the calculation of the \ac{BP} decoding thresholds of \ac{SC} \ac{MN} code ensembles over the \ac{biAWGN} channel, we resort to the \ac{EPC} model, with a \ac{BSC} as a-priori channel with crossover probability given by the \ac{DM} parameter $\omega$ and a \ac{biAWGN} communication channel. This allows mapping the \ac{BP} decoding thresholds of protograph-based \ac{SC} \ac{LDPC} code ensembles over the \ac{BSC}-\ac{biAWGN} parallel channels (Section~\ref{subsubsec:BSCAWGN}) to the thresholds of the corresponding protograph-based \ac{SC} \ac{MN} code ensembles, over the \ac{biAWGN} channel.
In particular, by fixing $\omega$, one fixes the rate of the \ac{SC} \ac{MN} code ensemble to $R = \Hb(\omega)$ as well as the conditional entropy of the a-priori channel in the \ac{EPC} model to $\ent{A} = \Hb(\omega)$. Hence, we have that the rate of the \ac{SC} \ac{MN} code ensemble equals the conditional entropy of the a-priori channel, i.e., $R = \ent{A}$. The entropy of the communication channel is $\ent{C} =  1-J(\sqrt{8E_s/N_0})$. It is then sufficient to translate the results of Fig. \ref{fig:thresholds_parallel} (rightmost chart) onto the rate-\ac{SNR} plane according the maps 
\begin{equation}
	R = \ent{A} \quad \text{and}\quad \frac{E_s}{N_0} = \frac{1}{8}\left[J^{-1}\left(1-\ent{C}\right)\right]^2.
\end{equation}
The result is depicted in Fig. \ref{fig:thresholds_cap}. Both the $(3,6)$  and the $(4,8)$ \ac{SC} \ac{MN} ensembles allow operating within $0.7$ dB from the \ac{biAWGN} channel capacity across all rates, by simply selecting the rate via the \ac{DM} parameter $\omega$. The $(4,8)$ ensemble, in particular, displays a maximum  gap of $\approx 0.15$ dB from the Shannon limit. The achieved thresholds largely outperform those attained by the block protograph \ac{MN} code ensembles found in \cite{zahr2024rate}, where the best protograph ensemble displayed a gap of about $1$ dB to the Shannon limit.

%%%%%%%%%%%%%%%%%%%%%%%%%%%%%%%%%%%%%%%%%%%%%%%%%%%%%%%%%%%%%%%%%%%%%%%%%%%%%%%%%%%%%%%%%%%%%%%%%%%%%%%%%%

%\section*{Acknowledgment}

%\bibliography{IEEEabrv,coding}

\end{document}